\documentclass{sig-alternate}

\usepackage{amstext}
\usepackage{amsfonts}

\conferenceinfo{STOC'02,}{May 19-21, 2002, Montreal, Quebec, Canada.} 
\CopyrightYear{2002} 
\crdata{1-58113-495-9/02/0005} 

\newcommand{\ignore}[1]{}

\newtheorem{theorem}{Theorem}[section]
\newtheorem{lemma}[theorem]{Lemma}
\newtheorem{corollary}[theorem]{Corollary}

\newtheorem{observation}[theorem]{Observation}

\newenvironment{remark}{{\sc Remark: }}{\qed}
\newenvironment{proofof}[1]{{\sc Proof of #1. }}{\qed}

\newcommand{\tensor}{\otimes}

\newcommand{\xor}{\oplus}

\newcommand{\meet}{\wedge}

\newcommand{\adjoint}{\dagger}

\newcommand{\trace}{{\rm Tr}}

\newcommand{\size}[1]{\left|#1\right|}

\newcommand{\ket}[1]{|#1\rangle}
\newcommand{\bra}[1]{\langle #1|}
\newcommand{\braket}[2]{\langle #1 | #2\rangle}
\newcommand{\ketbra}[2]{\ket{#1}\!\bra{#2}}
\newcommand{\density}[1]{\ketbra{#1}{#1}}
\newcommand{\norm}[1]{\left\|\,#1\,\right\|}

\newcommand{\set}[1]{{\left\{#1\right\}}}

\newcommand{\transpose}{{\mathsf T}}
\newcommand{\Span}{{\mathrm{span}}}

\newcommand{\pee}{{\mathcal{P}}}
\newcommand{\ip}{{\mathrm{IP}}}
\newcommand{\aitch}{{\mathcal{H}}}
\newcommand{\kay}{{\mathcal{K}}}

\begin{document}

\title{On communication over an entanglement-assisted
           quantum channel}

\numberofauthors{2}

\author{
\alignauthor Ashwin Nayak \titlenote{ 
  Supported by Charles Lee Powell Foundation, and NSF grants
  CCR~0049092 and EIA~0086038.
}\\
    \affaddr{ Computer Science Department, and}\\
    \affaddr{ Institute for Quantum Information}\\ 
    \affaddr{ California Institute of Technology}\\
    \affaddr{ Pasadena, CA~91125-8000}\\ 
    \email{ nayak@cs.caltech.edu}
\alignauthor Julia Salzman \titlenote{ 
  A part of this work was done while this author was visiting Caltech
  on a Summer Undergraduate Research Fellowship.
}\\
    \affaddr{ Mathematics Department}\\
    \affaddr{ Princeton University}\\
    \affaddr{ Fine Hall, Washington Road}\\
    \affaddr{ Princeton, NJ~08544-1000}\\
    \email{ jsalzman@princeton.edu}
}

\maketitle

\begin{abstract}
Shared entanglement is a resource available to parties communicating
over a quantum channel, much akin to public coins in classical
communication protocols. Whereas shared randomness does not  help in
the transmission of information, or significantly reduce the classical
complexity of computing functions (as compared to private-coin
protocols), shared entanglement leads to startling phenomena such as
``quantum teleportation'' and ``superdense coding.''

The problem of characterising the power of prior entanglement has
puzzled many researchers. In this paper, we revisit the problem of
transmitting classical bits over an entanglement-assisted quantum
channel. We derive a new, optimal bound on the number of quantum bits
required for this task, for any given probability of error. All known
lower bounds in the setting of bounded error entanglement-assisted
communication are based on sophisticated
information theoretic arguments. In contrast, our result is derived
from first principles, using a simple linear algebraic technique.
\end{abstract}




\section{Introduction}

Consider two parties solving a distributed task by communicating with
each other. Remarkably, it has been shown that if the two parties are
equipped with a quantum computer and can communicate by exchanging
quantum states, they can solve certain tasks at a significantly
smaller communication cost, when compared to classical
protocols~\cite{BuhrmanCW98, AmbainisSTVW98, Raz99}. This is
especially surprising since an early result due to
Holevo~\cite{Holevo73} (later explained in simpler terms by
Nayak~\cite{Nayak99}) rules out obvious methods of compressing
classical information into succinct quantum messages---Holevo's
theorem implies that~$n$ quantum bits of communication are necessary
to transmit~$n$ classical bits of information.

An additional resource that is available to parties communicating over
a quantum channel is ``shared entanglement'': the two parties may be
given some number of quantum bits jointly prepared in a fixed
superposition, prior to communicating with each other. For example,
they may jointly hold some number of EPR pairs.\footnote{An EPR pair
  consists of two qubits prepared in the maximally entangled
  state~$\frac{1}{\sqrt{2}} (\ket{00} + \ket{11})$.} The quantum
channel is then said to be ``entanglement-assisted.''

Shared randomness does not help in the transmission of information
from one party to another, or significantly reduce the classical
complexity of computing functions vis-a-vis private-coin
protocols~\cite[Section~3.3]{KushilevitzN97}. On the other hand, prior
entanglement leads to startling phenomena such as ``quantum
teleportation''~\cite{BennettBC+93} and ``superdense
coding''~\cite{BennettW92}. In particular, superdense coding allows us
to transmit~$n$ classical bits with perfect fidelity by sending
only~$n/2$ quantum bits. The problem of characterising the power of
prior entanglement has baffled many
researchers~\cite{BuhrmanW01,Klauck01}, especially in the setting of
bounded-error protocols. It is open whether it leads to more than a
factor of two savings (using superdense coding) or more than an
additive~$O(\log n)$ savings (when used to create shared randomness).
Few lower bounds are known for communication problems in this
setting~\cite{CleveDNT98,Nayak99b,Klauck00,KlauckNTZ01}, and are all
derived using sophisticated information-theoretic techniques.

In this paper, we focus on the most basic problem in the setting of
communication over an entanglement-assisted quantum channel, that of
transmitting classical bits from one party to another. We derive optimal
bounds on the number of quantum bits required for this task, for any
given probability of error. 

\begin{theorem}
\label{thm-main}
Suppose one party, Alice, wishes to communicate~$n$ bits to the other,
Bob, over an entanglement-assisted quantum channel. For any choice of
the entangled state, and any protocol such that the total number of
qubits sent by Alice to Bob (over all the rounds of communication)
is~$m_A$, let~$Y$ be the random variable
denoting Bob's output, when Alice wishes to convey~$X$.
If~$X$ is distributed uniformly
over~$\set{0,1}^n$, the probability that Bob correctly
recovers~$X$ is is bounded as
$$
\Pr[Y = X] ~~\le~~ \frac{2^{2m_A}}{2^n},
$$
irrespective of the number of qubits sent by Bob to Alice, or the number
of rounds of communication.
\end{theorem}
Thus, for protocols with probability of success~$\delta > 0$, we get
the optimal lower bound of~$m_A \ge {\frac{1}{2}}
(n-\log\frac{1}{\delta})$. This improves over the lower bound of~$m_A
\ge \frac{1}{2} (\delta n - H(\delta))$ implied by a result of Cleve
{\em et al.\/}~\cite[Theorem~2]{CleveDNT98}, when combined with Fano's
inequality~\cite[Section~2.11]{CoverT91}.

\begin{remark} 
  A bound for non-uniform distributions over the inputs also follows
  from the proof of Theorem~\ref{thm-main}, as in~\cite{Nayak99}.
  Note also that an optimal bound of~$n - \log\frac{1}{\delta}$ for
  the {\em total\/} number of quantum bits exchanged, including the
  communication required to create the prior entanglement, is implied
  by~\cite[Theorem~2.4]{Nayak99}.
\end{remark}

All known lower bounds for bounded-error
communication using prior entanglement
are based on complex information-theoretic arguments. In fact, one
might be lead to believe that such techniques are inevitable---any
lower bound proof necessarily depends on the property that the prior
shared state contains no information about the inputs.  Contrary to
this, our results are derived from first principles, using a linear
algebraic technique that has its roots in the work of
Nayak~\cite{Nayak99}. In order to prove Theorem~\ref{thm-main}, we
give a new characterisation of the joint state at the end of a quantum
protocol that complements the
characterisation due to Yao~\cite{Yao93}. It greatly clarifies the
role of shared entanglement in communication, and we expect that
it will further enhance our conceptual understanding of quantum
communication.

Putting Theorem~\ref{thm-main} together with a reduction due to Cleve
{\em et al.\/}~\cite{CleveDNT98}, we get a new lower bound
of~$\frac{1}{2} (n - 2\log\frac{1}{1-2\epsilon})$ for the
$\epsilon$-error entanglement-assisted communication complexity of the
inner product function.  The previous best lower bound
was~$\frac{1}{2} ((1-2\epsilon)^2 n - 1)$ due to~\cite{CleveDNT98}.
Since there is a classical~$n - \log\frac{1}{1-2\epsilon} + 1$ bit
public-coin protocol for Inner Product, and hence a~$\frac{1}{2} (n -
\log\frac{1}{1-2\epsilon} + 1)$ qubit quantum protocol with shared EPR
pairs, our lower bound is near-optimal.  Our results thus provide more
examples where shared entanglement leads to at most a factor of two
savings in communication.

The lower bound of~$\frac{1}{2} (n - 2\log\frac{1}{1-2\epsilon})$ for
Inner Product stated above was independently discovered by van Dam and
Hayden~\cite{DamH01} in the case of communication with shared {\em EPR
  pairs\/}.  However, they follow an information-theoretic approach
that provably breaks down in the presence of arbitrary prior
entanglement.

\subsection{Organisation of the paper}

The quantum communication model, and the associated terminology and
notation are described in Section~\ref{sec-prelims}. We begin by
analysing quantum {\em encoding\/} of classical bits in the presence
of entanglement in Section~\ref{sec-encoding}. In fact, we first
consider a very restricted kind of encoding, where the shared state
consists of EPR pairs, and no ancillary qubits are used in the
encoding (Section~\ref{sec-epr}). This contains the basic elements of
the proof for general encoding as well, which is the subject of
Section~\ref{sec-general}. Building on the insight gained from the
study of quantum encoding, we extend our results to the case of
interactive communication in Section~\ref{sec-inter}.

\section{Preliminaries}
\label{sec-prelims}

\subsection{The communication model}

In the quantum communication model of Yao~\cite{Yao93}, two parties
Alice and Bob hold qubits. When the game starts Alice holds a
superposition~$\ket{x}$ and Bob holds~$\ket{y}$, representing the
input to the two players. The initial joint state is thus~$\ket{x}_A
\tensor \ket{y}_B$, where a subscript indicates the player holding
that set of qubits. Furthermore each player has an arbitrarily large
supply of private qubits in some fixed basis state,
say~$\ket{\bar{0}}$.  The two parties then play in turns. Suppose it
is Alice's turn to play.  Alice can do an arbitrary unitary
transformation on her qubits and then send one or more qubits to Bob.
Sending qubits does not change the overall superposition, but rather
changes the ownership of the qubits, allowing Bob to apply his next
unitary transformation on the newly received qubits. At the end of the
protocol, one player measures one or more qubits in some basis, and
declares those as the result of the protocol. (In cases where a
specific player is required to know the answer, that player makes the
measurement.) In a classical probabilistic protocol the players may
only exchange messages composed of classical bits.

Note that there is no loss of generality in not allowing the players
to measure a subset of their quantum bits in the intermediate steps of
a protocol. This is because all measurements may be postponed to the
end by the {\em principle of safe storage\/}~\cite{BernsteinV97}. We
also assume, w.l.o.g., that the players do not modify the state of the
qubits containing their inputs.

In the classical model we can also define a {\em public-coin\/}
version, in which the players are also allowed to access a shared
source of random bits without any communication cost.  The classical
public and private-coin models are strongly related
(see~\cite[Section~3.3]{KushilevitzN97}).  In the quantum analogue of
the public-coin model, Alice and Bob may initially share an arbitrary
number of quantum bits which are in some pure state that is
independent of the inputs.  This is known as {\em communication with
  prior entanglement\/}~\cite{CleveDNT98, BuhrmanW01}, or in
information-theoretic terms, as {\em communication over an
  entanglement-assisted quantum channel\/}~\cite{BennettS98}.

The complexity of a quantum (or classical) protocol is the number of
qubits (respectively, bits) exchanged between the two players.  We say
a protocol {\em computes\/} a function~$f : X \times Y \mapsto
\{0,1\}$ with~$\epsilon \ge 0$ error if, for any input~$x \in X,y \in
Y$, the probability that the two players compute~$f(x,y)$ is at
least~$1-\epsilon$.  $Q_\epsilon(f)$ (resp.~$R_\epsilon(f)$) denotes
the complexity of the best quantum (resp.~probabilistic) protocol that
computes~$f$ with at most~$\epsilon$ error.  We will use the
notation~$Q^*_\epsilon(f)$ for entanglement-assisted quantum
communication of the function~$f$.

On occasion, we will concentrate on communication in one round, since
this often sheds light on fundamental properties of protocols for
certain problems. The message in a one-round protocol in which only
one player gets an input is called an {\em encoding\/} of the
input. The operations done by the other player, and her measurement
are together referred to as {\em decoding\/}.

\subsection{Miscellanea}

A {\em mixed state\/} over a set of qubits is a probability
distribution~$\set{p_i, \ket{\phi_i}}$ over superpositions (or {\em
  pure\/} states), where the state~$\ket{\phi_i}$ occurs with
probability~$p_i$.  We will sometimes use the notation~$\set{
  \ket{\phi_i}}$ for a mixed state, where the states~$\ket{\phi_i}$
are in general unnormalised, and are such that~$\sum_i \norm{\phi_i}^2
= 1$.

The following theorem gives a useful characterisation
of bi-partite quantum states (see~\cite[Section~2.5]{NielsenC00}).

\begin{theorem}[Schmidt decomposition theorem]
\label{thm-schmidt}
Any unit vector~$\ket{\phi}$ in a bi-partite Hilbert space~$\aitch
\tensor \kay$ may be represented as
$$
\ket{\phi} ~~=~~ \sum_i \sqrt{\lambda_i}\; \ket{e_i}\ket{f_i},
$$
where~$\set{\ket{e_i}}$ and~$\set{\ket{f_i}}$ are orthonormal sets
of states in~$\aitch$ and~$\kay$ respectively, and the~$\lambda_i \ge
0$ are such that~$\sum_i \lambda_i = 1$.
\end{theorem}

We denote the identity operator on states over~$k$ qubits by~$I_k$.

\section{Bounds for encoding}
\label{sec-encoding}

In this section we concentrate on one-way protocols, or {\em
  encoding\/}, by which one party, Alice, wishes to send some number
of classical bits to Bob.

\subsection{Encoding over EPR pairs, without ancilla}
\label{sec-epr}

We first prove our results in the case where Alice does not use any
ancillary qubits in the encoding process, and Alice and Bob share some
number of EPR pairs. This motivates the proof in the more general
case, and illustrates its essential elements.

We start with a simple property of maximally entangled states, such as
EPR pairs. This allows us to analyse the encoding process easily.
\begin{lemma}
\label{thm-epr}
For any unitary transformation~$U$ on~$E$ qu-bits, and any orthonormal
set~$\set{\ket{\phi_a} : a \in \set{0,1}^E }$ over~$E' \ge E$ qubits,
\begin{eqnarray*} 
\sum_{a \in \{0,1\}^{E}} U\ket{a} \ket{\phi_a}
  & = & \sum_{a \in \{0,1\}^{E}}
        \ket{a} \tilde{U} \ket{\phi_a},
\end{eqnarray*}
where~$\tilde{U}$ is any transformation on~$E'$ qubits such
that for all~$a,a' \in \set{0,1}^E$,
$\bra{\phi_a}\tilde{U}\ket{\phi_{a'}} = \bra{a'} U \ket{a}$.
\end{lemma}
\begin{proof}
Observe that for $b,c \in \{0,1 \}^{E}$, 
\begin{eqnarray*}
\bra{b}\bra{\phi_c} \sum_{a \in \{0,1\}^{E}} U\ket{a} \ket{\phi_a}
  & = & \bra{b}U\ket{c} \\
  & = & \bra{\phi_c} \tilde{U} \ket{\phi_b} \\
  & = & \bra{b}\bra{\phi_c} \sum_{a \in \{0,1\}^{E}} \ket{a}
  \tilde{U} \ket{\phi_a}.
\end{eqnarray*}
The lemma follows.
\end{proof}

We can now characterise the encoding process (without ancilla)
as follows.
\begin{lemma}
\label{thm-encoding1}
Suppose that Alice performs a unitary transformation on her share
of~$E$ EPR pairs, and then sends~$m$ of the~$E$ qubits to Bob. Then,
Bob has~$E+m$ qubits in a mixed state that can be represented
as~$\{p_{l}, |\phi_{l}\rangle \}$ ($l \in \{0,1\}^{E-m}$) with~$\{|
\phi_{l} \rangle \}_l$ orthonormal, and $p_{l}=\frac{1}{2^{E-m}}$.
\end{lemma}
\begin{proof}
Suppose that Alice applies a transformation~$V$ to her part of the
state. By Lemma~\ref{thm-epr}, the resulting state is
$$
\frac{1}{2^{E/2}} \sum_{a \in \{0,1\}^{E}} V \ket{a}_A \ket{a}_B
~~=~~ \frac{1}{2^{E/2}} \sum_{a \in \{0,1\}^{E}} \ket{a}_A
      {V^{\transpose}}\ket{a}_B. 
$$
After the communication, Alice and Bob's joint state may be
written as (w.l.o.g., Alice sends the rightmost~$m$ qubits to Bob):
$$
\frac{1}{2^{(E-m)/2}} \sum_{l \in \{0,1\}^{E-m}} \ket{l}_A ~~
\frac{1}{2^{m/2}} \sum_{r \in \set{0,1}^m } \ket{r}_B V^\transpose
\ket{lr}_B.
$$
Consider the mixed state on Bob's side obtained when Alice
measures her qubits in the standard basis. The probability~$p_{l}$
of Alice observing any given~$l$ is~$\frac{1}{2^{E-m}}$. The
state of Bob's~$E+m$ qubits when Alice gets outcome~$l$ is
$$
\ket{\phi_{l}} ~~=~~ \frac{1}{2^{m/2}} \sum_{r \in \{0,1\}^{m}}
\ket{r} V^\transpose \ket{lr}.
$$
We may easily verify that these are orthonormal for different~$l$:
\begin{eqnarray*}
\braket{\phi_{l}}{\phi_{l'}}
  & = & \frac{1}{2^m} \sum_r \bra{lr} V^* V^\transpose \ket{l'r} \\
  & = & \frac{1}{2^m} \sum_r \braket{lr}{l'r} \\
  & = & \delta_{l,l'}. \\
\end{eqnarray*}
Note that the above measurement by Alice does not affect the decoding
process; Bob's density matrix remains unchanged by it
(see~\cite[Section~2.4]{NielsenC00}, especially Section~2.4.3).
Nonetheless, it allows us to express Bob's mixed state in a convenient
form.

This proves the lemma.
\end{proof}

By a simple dimensional argument, can now get an alternative proof of
the fact that the superdense coding scheme of~\cite{BennettW92} is
optimal (in the case of encoding without ancilla). We omit the proof.
\ignore{
\begin{theorem}
\label{thm-noerror1}
If Alice encodes messages over her part of~$E$ shared EPR pairs and
sends~$m \le E$ qubits to Bob, she can communicate at most~$2m$
classical bits without error.
\end{theorem}
\begin{proof}
Suppose there is an encoding scheme with~$E$ EPR pairs and~$m$ bits of
communication whereby Alice is is able to
transmit~$n$ bits to Bob without error. Let~$\set{p_{x,l},
\ket{\phi_{x,l}} }_l$ be Bob's mixed state when Alice transmits~$x
\in \set{0,1}^n$, as given by Lemma~\ref{thm-encoding1},
and let~$H_x = \Span \set{ \ket{\phi_{x,l}} }_l$. 
Then~$\dim H_x = 2^{E-m}$.

If we require Bob to decode Alice's messages with perfect fidelity, we
have~$H_x \perp H_y$, if $x \neq y$.  It follows that
\begin{equation}
\label{eqn-union} 
\dim  \bigoplus_{x \in \set{0,1}^n } {H_x} ~~=~~ 2^{n} \cdot 2^{E-m}.
\end{equation}
On the other hand, since the code words are over~$E+m$ qubits,  
\begin{equation} 
\label{eqn-dimbound}
\dim \bigoplus_{x \in \set{0,1}^n} {H_x} ~~\leq~~ 2^{E+m}.
\end{equation}
Equations~(\ref{eqn-union}) and~(\ref{eqn-dimbound}) imply that~$n
 \leq 2m$, as desired.
\end{proof}
}

In general, we can tolerate a little error in the decoding process.
This opens up the possibility of Alice being able to reduce the
communication significantly. The following theorem places limits on
the savings achieved.
\begin{theorem}
\label{thm-errorbd1}
If Alice encodes messages~$x \in \set{0,1}^n$ over EPR pairs without
ancilla, and sends~$m$ qubits to Bob, the probability of correct
decoding of a message chosen uniformly at random is bounded
as~$\Pr[\textrm{correct decoding}] \leq \frac{2^{2m}}{2^n}$. 
\end{theorem}
\begin{proof}
Suppose that the number of EPR pairs Alice and Bob share initially
is~$E$.  Let~$\set{p_{x,l}, \ket{\phi_{x,l}} }_l$ be Bob's mixed
state when Alice has input~$x \in \set{0,1}^n$, as given by
Lemma~\ref{thm-encoding1}.

We may view the entire decoding procedure used by Bob as measuring the
encoded state with some ancillary qubits (w.l.o.g., assumed to be
initialised to state~$\ket{\bar{0}}$) with the projection
operators~$\set{P_y}$. Here, the outcome~$y \in \set{0,1}^n$
corresponds to Bob's guess for the encoded message.  We will omit the
ancilla from the expressions below, for clarity of exposition.

Let $C$ be the event that Bob decodes a message correctly, $C_x$
($C_{x,l}$) that he does so on receiving the encoding of~$x$
($\ket{\phi_{x,l}}$, respectively).  Let~$x$ be the event that Alice
encodes message~$x$, and~$x_l$ that~$\ket{\phi_{x,l}}$ is prepared
given that~$x$ is encoded.  Then
\begin{eqnarray}
\Pr[C] 
  & = & \sum_x \Pr[C_x] \cdot \Pr[x] \nonumber \\
  & = & \sum_{x} \frac {\Pr[C_x]}{2^n} \nonumber \\
  & = & \sum_{x,l} \frac{\Pr[C_{x,l}] \cdot \Pr[x_l]}{2^n} \nonumber \\
\label{eqn-pC}
  & = & \sum_{x,l} \frac{ \Pr[C_{x,l}] }{2^{E-m}2^n}.
\end{eqnarray}
It thus suffices to bound~$\sum_{x,l} \Pr[C_{x,l}] $.  Observe that
\begin{equation} 
\label{eqn-pCxl}
\Pr[C_{x,l}] ~~=~~ \norm{P_x \ket{\phi_{x,l}}}^2.
\end{equation}

We introduce some notation.  For each~$x$, let~$H_x$ be the space
spanned by~$\set{ \ket{\phi_{x,l}} }_l$. Note that~$\set{
  \ket{\phi_{x,l}} }_l$ is an orthonormal basis for~$H_x$. Let~$R_x$
be the projection onto~$H_x$.  Since we allow a little error in the
decoding process, the different spaces~$H_x$ may not be orthogonal.

Let~$H$ be the space spanned by all the vectors~$\set{
  \ket{\phi_{x,l}} }_{x,l}$, and~$Q$ the projection operator onto~$H$.
For each~$x$, let the set~$\set{ \ket{e_{x,j}}}_j$ be an orthonormal
basis for the range of~$P_x$.  Then~$\set{ \ket{e_{x,j}}}_{x,j}$ is an
orthonormal basis for the entire decoding space.

Now,
\begin{eqnarray}
\sum_l \norm{ P_x \ket{\phi_{x,l}} }^2
  & = & \sum_{l,j} \size{ \braket{e_{x,j}}{\phi_{x,l}} }^2 \nonumber \\
  & = & \sum_{j} \norm{ R_x \ket{e_{x,j}} }^2 \nonumber \\
\label{eqn-pphi1}
  & \le & \sum_j \norm{ Q \ket{e_{x,j}} }^2,
\end{eqnarray}
since the length of the projection of~$\ket{e_{x,j}}$ onto~$H_x$ is at
most the length of its projection on the space~$H$ (of which~$H_x$ is
a subspace).

From equation~(\ref{eqn-pphi1}), 
\begin{eqnarray}
\sum_{x,l} \norm{ P_x \ket{\phi_{x,l}} }^2
  & \le &  \sum_{x,j} \norm{ Q \ket{e_{x,j}} }^2 \nonumber \\
  & = &    \sum_{x,j} \bra{e_{x,j}} Q \ket{e_{x,j}} \nonumber \\
  & = &    \trace \; Q ~~=~~ \dim H \nonumber \\
\label{eqn-pphi2}
  & \le &  2^{E+m},
\end{eqnarray}
since the space~$H$ is generated by states over~$E+m$ qubits.

Combining equations~(\ref{eqn-pC}), (\ref{eqn-pCxl}), and~(\ref{eqn-pphi2}),
we get
$$
\Pr[C] ~~\leq~~ \frac{2^{E+m}}{2^{E-m} 2^n} ~~=~~ \frac{2^{2m}}{2^n},
$$
as claimed.
\end{proof}

Encoding with EPR pairs and ancilla leads to states very similar to
those in Lemma~\ref{thm-encoding1}, and Theorem~\ref{thm-errorbd1}
holds in that case as well. We will however skip ahead to encoding
where Alice uses extra space, and an arbitrary entangled state.

\subsection{Encoding with general prior entanglement}
\label{sec-general}

In general, in trying to transmit information, Alice and Bob may share
an {\em arbitrary\/} entangled state (independent of their inputs) 
before they
interact. In this section we show that the results in the previous
section apply irrespective of which initial entangled state Alice and Bob
share.

The main difficulty here is that the property of messages encoded
over EPR pairs embodied in Lemma~\ref{thm-encoding1} may fail to hold.
However, we show a simple connection between encoding with EPR pairs
and encoding with an arbitrary entangled state that allows us to
conclude an identical result.

We start by observing that we need only consider protocols which make
use of a special kind of shared state.
\begin{observation}
\label{thm-state} 
In any quantum communication protocol with prior entanglement, we may
assume, without loss of generality, that the initial shared state is
of the form
$$ 
\sum_{a \in \set{0,1}^E} \sqrt{\lambda_a}\; \ket{a}_A
\ket{a}_B,
$$
where~$\lambda_a$ are non-negative reals, and~$\sum_a
\lambda_a = 1$.
\end{observation}
\begin{proof}
This follows directly from the Schmidt decomposition theorem
(Theorem~\ref{thm-schmidt}).
Consider a protocol~$\pee$ in which
the quantum state shared by Alice and Bob has~$E_A$
qubits on Alice's side and~$E_B$ qubits on Bob's side. For
concreteness, assume that~$E_A \le E_B$.  By
Theorem~\ref{thm-schmidt}, the shared state may be
expressed as
$$
\sum_{b \in \set{0,1}^{E_A}} \sqrt{\mu_b}\; \ket{\phi_b}_A
\ket{\psi_b}_B,
$$
where the~$\mu_b$ are non-negative reals summing up to~$1$, and the
sets~$\set{\ket{\phi_b}}$ and~$\set{\ket{\psi_b}}$ are orthonormal. We
may modify the protocol to a new protocol~$\pee'$, which has the same
behaviour as~$\pee$ on each input, but where the shared state is of
the form in stated in the observation above. Consider any unitary
transformations~$U,V$ on~$E = E_B$ qubits such that for every~$b \in
\set{0,1}^{E_A}$,
\begin{eqnarray*}
U & : & \ket{\bar{0},b} \mapsto \ket{\bar{0}}\ket{\phi_b} \\
V & : & \ket{\bar{0},b} \mapsto \ket{\psi_b}.
\end{eqnarray*}
Let~$\lambda_{\bar{0}b} = \mu_b$, for~$b$ as above, and let the rest
of the~$\lambda_a$ be~$0$.  The protocol~$\pee'$ begins with the
shared state
$$
\sum_{a \in \set{0,1}^E} \sqrt{\lambda_a}\; \ket{a}_A
\ket{a}_B,
$$
and then Alice and Bob apply~$U$ and~$V$ to their qubits
respectively. Thereafter, the protocol proceeds exactly as in~$\pee$.
By construction, the protocols behave the same way for each
input.
\end{proof}

We make another simplifying observation about the protocols that we
need consider.
\begin{observation}
\label{thm-ancilla}
In any quantum communication protocol with prior entanglement, we may
assume, without loss of generality, that neither Alice nor Bob uses any
ancillary qubits in their local unitary operations or measurements.
\end{observation}
This is because all the ancillary qubits used may be considered as part
of the initial shared state.

The above observations allow us to relate the encoding with a general
entangled state to the encoding obtained when EPR pairs are used
instead.
\begin{lemma}
\label{thm-encoding3}
Suppose that Alice performs a unitary transformation on her share of
the joint state
$$
\sum_{a \in \set{0,1}^E} \sqrt{\lambda_a}\;
\ket{a}_A \ket{a}_B,
$$
and then sends~$m$ of the~$E$ qubits to Bob.  Then, Bob has~$E+m$
qubits in a mixed state that can be represented as
$$
\set{ 2^{m/2}
  (I_m \tensor \Lambda) \ket{\phi_{l}} }_{l \in \{0,1\}^{E-m}}
$$
with~$\set{ \ket{\phi_{l}} }_l$ orthonormal, and~$\Lambda = \sum_a
\sqrt{\lambda_a}\; \density{a}$.
\end{lemma}
\begin{proof}
Note that the shared state may be written as
$$
(I_E \tensor \Lambda) \sum_a \ket{a} \ket{a}.
$$
Suppose Alice applies the transformation~$V$ to her~$E$ qubits. The
resulting joint state is
\begin{eqnarray}
\lefteqn{ (V \tensor I_E)(I_E \tensor \Lambda) \sum_a \ket{a} \ket{a}
  } \nonumber \\
  & = & (I_E \tensor \Lambda) \sum_a V \ket{a} \ket{a} \nonumber \\
\label{eqn-2nd}
  & = & (I_E \tensor \Lambda) \sum_a \ket{a} V^\transpose \ket{a} \\
  & = & \sum_{l \in \set{0,1}^{E-m}} \ket{l} \;\; (I_m \tensor \Lambda)
        \sum_{r \in \set{0,1}^m} \ket{r} V^\transpose \ket{lr},
        \nonumber
\end{eqnarray}
where equation~(\ref{eqn-2nd}) follows from Lemma~\ref{thm-epr}. Let, as
in Lemma~\ref{thm-encoding1},
$$
\ket{\phi_l} ~~=~~ 2^{-m/2} \sum_{r \in \set{0,1}^m} \ket{r}
V^\transpose \ket{lr}.
$$
Suppose Alice sends~$m$ of her qubits to Bob, and measures the remaining
qubits in the standard basis. The residual, unnormalised, state with
Bob is then~$2^{m/2} (I_m \tensor \Lambda) \ket{\phi_l}$, when she
observes~$l \in \set{0,1}^{E-m}$. That the states~$\ket{\phi_l}$ are
orthonormal is shown in the proof of Lemma~\ref{thm-encoding1}.
\end{proof}

We can now prove the equivalent of Theorem~\ref{thm-errorbd1} when
Alice and Bob share an arbitrary entangled state.
\begin{theorem}
\label{thm-errorbd3}
If Alice encodes~$2^n$ messages over her part of an arbitrary (but
fixed) shared entangled state and
some ancillary qubits, and sends~$m$ qubits to Bob, the probability of
correct decoding of a message chosen uniformly at random is bounded
as~$\Pr[\textrm{correct decoding}] \leq \frac{2^{2m}}{2^n}$.
\end{theorem}
\begin{proof}
We use the same notation as in the proof of Theorem~\ref{thm-errorbd1},
adapted to the different encoding we get here due to the more
general entangled state.

By Observation~\ref{thm-ancilla}, we may assume that Alice and Bob
operate only on their shared entangled state. We may further assume
that this state is of the special form described in
Observation~\ref{thm-state}.
 
Let~$\set{ 2^{m/2} (I_m \tensor \Lambda) \ket{\phi_{x,l}} }_l$ be
Bob's mixed state when Alice encodes~$x \in \set{0,1}^n$, as given by
Lemma~\ref{thm-encoding3}. Since no ancilla is used in the decoding
procedure (i.e., in Bob's measurement to extract~$x$, cf.\
Observation~\ref{thm-ancilla}), the projection operators~$P_y$ are
over~$E+m$ qubits.
Now,
\begin{eqnarray}
\label{eqn-pC3}
\Pr[C] 
  & = & \sum_{x} \frac{\Pr[C_{x}]}{2^n},
             \textrm{~~~~and} \\
\label{eqn-pCxl3}
\Pr[C_{x}]
  & = & 2^m \sum_l \norm{ P_x (I_m \tensor \Lambda) \ket{\phi_{x,l}} }^2.
\end{eqnarray}
Furthermore,
\begin{eqnarray}
\sum_{x,l} \norm{ P_x (I_m \tensor \Lambda) \ket{\phi_{x,l}} }^2 
  & = & \sum_{x,l,j} \size{ \bra{e_{x,j}} (I_m \tensor \Lambda)
        \ket{\phi_{x,l}} }^2 \nonumber \\
  & = & \sum_{x,j} \norm{ R_x (I_m \tensor \Lambda) \ket{e_{x,j}} }^2
        \nonumber \\ 
  & \le & \sum_{x,j} \norm{(I_m \tensor \Lambda) \ket{e_{x,j}} }^2
          \nonumber \\
  & = & \sum_{x,j} \bra{e_{x,j}} (I_m \tensor \Lambda^2) \ket{e_{x,j}}
        \nonumber \\
  & = & \trace\;\; (I_m \tensor \Lambda^2) \nonumber \\
\label{eqn-pphi3}
  & = & 2^m \sum_a \lambda_a ~~=~~ 2^m.
\end{eqnarray}
Combining equations~(\ref{eqn-pC3}), (\ref{eqn-pCxl3})
and~(\ref{eqn-pphi3}), we get
$$\Pr[C] \le 2^{2m}/2^n.$$
\end{proof}

\section{Extension to interactive \\ communication}
\label{sec-inter}

\subsection{The main lemma}

In this section, we analyse the most general quantum protocols for
exchanging information. In these protocols, Alice and Bob share an
arbitrary entangled state to begin with, and exchange messages both
ways in order to communicate.

The essential idea behind the results below is contained in
Lemma~\ref{thm-encoding3}, and leads to a new characterisation of the
joint state in quantum protocols. In order to prove the lemma from
first principles, we focus on protocols in which there is no prior
entanglement. That it holds also for communication with prior
entanglement may be inferred from the lemma itself by applying it to a
protocol in which the prior shared entanglement is generated by Bob
creating the state to be shared, and sending the appropriate part of
it to Alice.

\begin{lemma}
\label{thm-encoding}
Let~$\pee$ be any quantum communication protocol (without prior
entanglement) in which the number of qubits sent by Alice to Bob (Bob
to Alice) is~$m_A$ (respectively,~$m_B$), and the final number of qubits with
Alice (Bob) is~$q_A$ (respectively,~$q_B$). Then, the joint state of Alice
and Bob at the end of the protocol may be expressed as $$
\sum_{a \in \set{0,1}^{q_A}} \ket{a}_A \; \Lambda \ket{\phi_a}_B,
$$
where
\begin{enumerate}
\item
$\Lambda$ is a linear transformation that maps~$q_B + 2m_B$ qubits
to~$q_B$ qubits, depends only on the unitary transformations of Bob,
and satisfies~$\trace(\Lambda\Lambda^\adjoint) = 2^{2m_A}$, and

\item
$\set{\ket{\phi_a}}$ is an orthonormal set of states over~$q_B + 2m_B$
qubits, and depends only on the unitary transformations
of Alice.

\end{enumerate}
\end{lemma}
\begin{proof}
The proof goes by induction on the number of rounds~$t$. 

In the beginning (for~$t = 0$), the joint state (w.l.o.g.)
is~$\ket{\bar{0}}_A \tensor \ket{\bar{0}}_B$, which represents {\em
  all\/} the qubits the two players use during the protocol. This is
of the form described in the lemma, with~$\Lambda = \Lambda_0 =
\density{\bar{0}}$.

Let~$q_{A,t}, q_{B,t},m_{A,t},m_{B,t}$ be the quantities corresponding
to~$q_A,q_B,m_A,m_B$ after~$t \ge 0$ rounds of communication. Assume
that at this stage, the joint state of Alice and Bob is
$$
\sum_{a \in \set{0,1}^{q_{A,t}}} \ket{a}_A \; \Lambda_t \ket{\phi_{a,t}}_B,
$$ 
where~$\Lambda_t$ and~$\set{\ket{\phi_{a,t}}}$ satisfy the
conditions stated in the lemma (in terms of~$q_{B,t},m_{A,t},m_{B,t}$).
We look at two cases for the~$(t+1)$'th round of communication.

{{\sc Case}~(a).~} Alice applies a unitary transformation~$U$ to her
qubits and sends~$p$ qubits to Bob.

The state after the unitary transformation is
\begin{eqnarray*}
\lefteqn{ (U \tensor I_{q_{B,t}}) (I_{q_{A,t}} \tensor \Lambda_t) 
    \sum_{a \in \set{0,1}^{q_{A,t}}} \ket{a} \ket{\phi_{a,t}} } \\
  & = & (I \tensor \Lambda_t) \sum_{a \in \set{0,1}^{q_{A,t}}}
        U\ket{a} \ket{\phi_{a,t}} \\
  & = & (I \tensor \Lambda_t) \sum_{a \in \set{0,1}^{q_{A,t}}} \ket{a} \;
        \tilde{U} \ket{\phi_{a,t}},
\end{eqnarray*}
where~$\tilde{U}$ is a unitary transformation on Bob's qubits as given
by Lemma~\ref{thm-epr}. Thus, after Alice sends~$p$ of
her qubits to Bob (w.l.o.g., these are the~$p$ rightmost qubits),
the joint state looks like
\begin{equation}
\label{eqn-a-state}
\sum_{l \in \set{0,1}^{q_{A,t+1}}} \ket{l}_A \; (I_p \tensor \Lambda_t)
\sum_{r \in \set{0,1}^p} \ket{r}_B \; \tilde{U}
\ket{\phi_{lr,t}}_B.
\end{equation}
Here, $q_{A,t+1} = q_{A,t} - p$, $q_{B,t+1} = q_{B,t} + p$, $m_{A,t+1}
= m_{A,t} + p$, and~$m_{B,t+1} = m_{B,t}$.

Let
\begin{eqnarray*}
\Lambda_{t+1} 
  & = & 2^{p/2} (I_p \tensor \Lambda_t), \textrm{~~~and} \\
\ket{\phi_{l,t+1}} 
  & = & 2^{-p/2} \sum_{r} \ket{r} \; \tilde{U}
  \ket{\phi_{lr,t}}. \\
\end{eqnarray*}
Now,
\begin{eqnarray*}
\trace\; \Lambda_{t+1}\Lambda_{t+1}^\adjoint 
  & = &  2^p\; \trace\; (I_p \tensor \Lambda_t \Lambda_t^\adjoint) \\
  & = &  2^p \cdot 2^p \cdot 2^{2m_{A,t}} \\
  & = &  2^{2m_{A,t+1}}.
\end{eqnarray*}
Moreover, for the same reasons as in the proof of
Lemma~\ref{thm-encoding1}, the set~$\set{\ket{\phi_{l,t+1}}}$ is
orthonormal. Thus, the state in equation~(\ref{eqn-a-state}) is of the
form stated in the lemma.

{{\sc Case}~(b).~} Bob applies a unitary transformation~$V$ to his
qubits and sends~$p$ qubits to Alice. W.l.o.g., these are the~$p$
leftmost qubits.

After the communication, the joint state looks like
$$
\sum_{a \in \set{0,1}^{q_{A,t}}} 
    \sum_{l \in \set{0,1}^p} \ket{a}_A \ket{l}_A \;\; 
    (\bra{l} \tensor I_{q_{B,t}-p}) V \Lambda_t \ket{\phi_{a,t}}_B,
$$
which may be recast as
\begin{equation}
\label{eqn-b-state}
  \sum_{a,l} \ket{al} \; \Lambda_{t+1} \ket{\phi_{al,t+1}},
\end{equation}
where
\begin{eqnarray*}
\Lambda_{t+1} 
  & = & \sum_{b \in \set{0,1}^p} (\bra{b} \tensor I_{q_{B,t} - p}) \; V
        \Lambda_t \; (\bra{b} \tensor I_{q_{B,t}}),  \\ 
\ket{\phi_{al,t+1}}
  & = & \ket{l}\ket{\phi_{a,t}}.
\end{eqnarray*}
Now $m_{A,t+1} = m_{A,t}$, $m_{B,t+1} = m_{B,t} + p$, $q_{A,t+1} =
q_{A,t} + p$, and~$q_{B,t+1} = q_{B,t} -p$.

The states~$\ket{\phi_{al,t+1}}$ are orthonormal. Moreover, 
\begin{eqnarray*}
\lefteqn{ \trace\; \Lambda_{t+1} \Lambda_{t+1}^\adjoint } \\
  & = & \trace \sum_{b,b'} (\bra{b} \tensor I) V \Lambda_t (\bra{b}
        \tensor I) (\ket{b'} \tensor I) \Lambda_t^\adjoint
        V^\adjoint (\ket{b} \tensor I) \\ 
  & = & \sum_b \trace \left[  (\bra{b} \tensor I) \; V \Lambda_t
        \Lambda_t^\adjoint V^\adjoint \; (\ket{b} \tensor I) \right] \\ 
  & = & \sum_b \trace \left[ (\density{b} \tensor I) \; V \Lambda_t
        \Lambda_t^\adjoint V^\adjoint \right] \\
  & = & \trace \left[ (I \tensor I) \; V \Lambda_t \Lambda_t^\adjoint
        V^\adjoint \right] \\ 
  & = & \trace\; \Lambda_t \Lambda_t^\adjoint ~~=~~ 2^{2 m_{A,t+1}}.
\end{eqnarray*}
Thus, the state in equation~(\ref{eqn-b-state}) is of the form
described in the lemma.

This completes the induction step, and the proof.
\end{proof}

\subsection{Implications for communication problems}

We now sketch how our characterisation of quantum protocols
enables us to prove Theorem~\ref{thm-main}.

\begin{proofof}{Theorem~\ref{thm-main}}
The proof is essentially the same as for Theorem~\ref{thm-errorbd3},
and we use the same notation here.

Lemma~\ref{thm-encoding} shows that Bob's state remains of
form similar to that
in Lemma~\ref{thm-encoding3} as he interacts with Alice
during the protocol.  Let~$\set{ \Lambda \ket{\phi_{x,l}} }_l$ be
Bob's mixed state at the end of the protocol when Alice has input~$x
\in \set{0,1}^n$, as given by Lemma~\ref{thm-encoding}. Note
that~$\Lambda$ is independent of~$x$.

Since we may assume that all the ancillary qubits used by Bob are
included in his state above (cf.\ the proof of
Lemma~\ref{thm-encoding}), the projection operators~$P_y$ are
over~$q_B$ qubits.  Now, as before,
$$
\Pr[C]  
  ~~=~~ \sum_{x} \frac{\Pr[C_{x}]}{2^n}  
  ~~=~~ \frac{1}{2^n} \sum_{x,l} \norm{ P_x \Lambda \ket{\phi_{x,l}} }^2.
$$
Furthermore, 
\begin{eqnarray*}
\sum_{x,l} \norm{ P_x \Lambda \ket{\phi_{x,l}} }^2
  & = &    \sum_{x,l,j} \size{ \bra{e_{x,j}}\Lambda \ket{\phi_{x,l}} }^2 \\
  & \le &  \sum_{x,j} \norm{\Lambda^\adjoint \ket{e_{x,j}} }^2 \\
  & = &    \trace\; \Lambda\Lambda^\adjoint ~~=~~ 2^{2m_A}.
\end{eqnarray*}
Combining these, we get~$\Pr[C] \le 2^{2m_A}/2^n$.
\end{proofof}

Finally, we apply Theorem~\ref{thm-main} to obtain an improved lower
bound for the entanglement-assisted quantum communication complexity
of the inner product function~$\ip_n$. (The function~$\ip_n :
\set{0,1}^n \times \set{0,1}^n \rightarrow \set{0,1}$ is defined
as~$\ip_n(x,y) = \xor_i (x_i \meet y_i)$.)  The connection between the
two is provided by the following reduction due to Cleve {\it et
  al.\/}~\cite{CleveDNT98}.

\begin{theorem}[Cleve, van~Dam, Nielsen, Tapp]
\label{thm-reduction}
If $Q^{*}_\epsilon(\ip_n) = m$, then there is an entanglement-assisted
protocol for transmitting~$n$ bits with probability of success
at least~$(1-2\epsilon)^2$, such that the total communication from each
party to the other, over all the rounds of communication, is~$m$ qubits.
\end{theorem}

Theorem~\ref{thm-main} now implies
\begin{corollary}
\label{thm-ip-lb}
$Q^{*}_\epsilon(\ip_n) ~~\ge~~ \frac{1}{2} (n -
\log\frac{1}{(1-2\epsilon)^2})$.
\end{corollary}

It is not hard to see that for any~$\epsilon < 1/2$, there is a
public-coin randomised protocol for~$\ip_n$ with communication cost at
most~$n-\log\frac{1}{1-2\epsilon} + 1$. Along with the superdense
coding scheme of~\cite{BennettW92}, this means that

\begin{theorem}
\label{thm-ip-ub}
$Q^{*}_\epsilon(\ip_n) ~~\le~~  \frac{1}{2}
(n-\log\frac{1}{1-2\epsilon} + 1)$.
\end{theorem}
Thus, our lower bound is close to optimal, and for constant error, is
within an additive~$O(1)$ term of the upper bound.
Since~$Q_{1/3}(\ip_n) \le n$, this provides more evidence
that prior entanglement does not give us a saving of more than a
factor~$2$ (plus perhaps an additive term of~$O(\log n)$) in
communication cost.

\section{Acknowledgements}

We would like to thank Leonard Schulman for insightful discussions,
and Umesh Vazirani and the anonymous referees for helpful comments on
the paper.



\end{document}